\documentclass[prd,aps,twocolumn,superscriptaddress,nofootinbib,amsmath,amssymb,aps]{revtex4-2}
\usepackage{bbold}
\usepackage{float}
\usepackage{hhline}
\usepackage[table]{xcolor}
\usepackage{array,multirow}

\usepackage[hidelinks]{hyperref}
\hypersetup{
    colorlinks=false,
    linkcolor=blue,
    filecolor=magenta,      
    urlcolor=cyan,
    pdftitle={Overleaf Example},
    pdfpagemode=FullScreen,
    }

\usepackage{xcolor}
\definecolor{purple}{rgb}{0.8,0,0.6}

\usepackage{graphicx}
\graphicspath{ {figures/} }
\usepackage{bm}
\usepackage{braket}
\usepackage{slashed}

\newcommand{\beqn}{\begin{eqnarray}}
\newcommand{\eeqn}{\end{eqnarray}}
\newcommand{\beqs}{\begin{subequations}}
\newcommand{\eeqs}{\end{subequations}\\[-3.5mm]\noindent}
\newcommand{\eq}[1]{(\ref{#1})}

\newcommand\blfootnote[1]{%
  \begingroup
  \renewcommand\thefootnote{}\footnote{#1}%
  \addtocounter{footnote}{-1}%
  \endgroup
}

\begin{document}

\title{Quark Gas at High Temperature: Finite-Volume Effects
}

\author{R.N. Rogalyov}
\email{rnr@ihep.ru}
\affiliation{NRC ``Kurchatov Institute'' - IHEP,
            Protvino, Moscow region, 142281, Russia}

\author{N.V. Gerasimeniuk}
\email{gerasimenyuk.nv@dvfu.ru}
\affiliation{Pacific Quantum Center, Far Eastern Federal University,
            Vladivostok, 690922, Russia}

\author{A.A. Korneev}
\email{korneev.aa@dvfu.ru}
\affiliation{Pacific Quantum Center, Far Eastern Federal University,
            Vladivostok, 690922, Russia}

\date{\today}

\begin{abstract}
It is shown that the textbook formula for the pressure of free massless fermions leads to negative values of the probability mass function of the distribution of the system of free massless fermions in the fermion number. The ways to resolve this paradox are proposed. A detailed analysis of the corresponding partition function indicates the presence of a Roberge-Weiss transition in the absence of strong interactions.
\end{abstract}

\maketitle

\section{Introduction}

\blfootnote{\textit{This is a preprint of the work accepted for publication in journal
"Physics of Particles and Nuclei" 2025 \copyright Pleiades Publishing,
Ltd. 2025 (\href{http://pleiades.online}{http://pleiades.online})}}
The baryon number distributions of fireballs produced in collisions of heavy nuclei have been intensively studied in recent years 
\cite{Braun-Munzinger:2011shf,Behera:2018wqk,Li:2023kja}. Such distributions carry important information about the dynamics of strongly interacting matter.
In particular, higher cumulants of such distributions could show an increase in fluctuations of the baryon number $n$, which indicates on the critical point in the $\mu_B-T$ phase plane of dense baryon matter~\cite{Bzdak:2019pkr}. Moreover, they could also provide information about the initial stage of fireball evolution.

The distribution of fireballs by baryon number is closely related to the dependence of the baryon density $\rho$ and the corresponding pressure $p$ on the baryon chemical potential $\mu_B$. Such dependence undergoes significant changes as the temperature increases from the pseudocritical chiral crossover temperature $T_\mathrm{pc}\approx 154$~MeV to the Roberge-Weiss temperature $T_\mathrm{RW}\approx 208$~MeV~\cite{Roberge:1986mm,Bonati2016,Begun_2021}.

By numerical simulation of lattice QCD at high temperatures ($T>T_{RW}$), it was shown~\cite{Bazavov:2017dus} that the first few coefficients of pressure expansion in the Taylor series in $\mu_B$ coincide with the corresponding coefficients of a free massless quark gas.
It was also demonstrated that~\cite{Bornyakov:2016wld} for $T>T_{RW}$, the dependence of the baryon number density on $\mathrm{Im}\,\mu_B$ is well fit by a polynomial of the third degree, which corresponds to a  free massless quark gas.

Knowing the baryon density and using numerical integration methods with great accuracy, one can find the probability $P_n$ that the baryon number of the system is equal to $n$. However, it turned out that for large $n$ probabilities $P_n$ take unphysical negative values. In \cite{Roberge:1986mm} and~\cite{Bornyakov:2022blw} this difficulty was overcome by calculating $P_n$ in the infinite-volume limit using approximate asymptotic formulas that give physically meaningful probability values. The study of the reasons for the discrepancy between the results obtained using numerical and asymptotic methods is the subject of this work.

We found that negative probabilities appear even in the case of free massless fermions when using quite traditional calculation methods presented in textbooks, e.g.~\cite{Greiner:1995,Kapusta:2006pm,Cleymans1986}. The paradox arises when the calculation of the pressure 
involves an incorrect transition from summation to integration. 
In this paper, we describe in detail the calculation of pressure and baryon number distribution for a free massless fermion gas. Since the described problem arises even in the 
one-dimensional case, we consider this case in detail.

We demonstrate that, in the one-dimensional case, the asymptotic formula can give a fairly good approximation to the exact result.  We also show that the position of the Lee-Yang zeros in both one-dimensional and three-dimensional case indicates the presence of the Roberge-Weiss transition in the free theory.

In the Section~\ref{sec:negative_probabilities} we formulate the problem of negative probabilities. Section~\ref{sec:1D} is devoted to a detailed comparison of the results obtained by the naive transition to the limit of infinite volume with exact and asymptotic formulas in the one-dimensional case. The Roberge-Weiss transition and the justification of the asymptotic formulas are discussed in Section~\ref{sec:RW}.

\section{The problem of negative probabilities}
\label{sec:negative_probabilities}

We consider a system with the Hamiltonian $\hat H$
and the baryon number operator $\hat B$ ($[\hat B, \hat H]=0$). The corresponding grand canonical partition function has the form
\begin{equation}\label{eq:ZGC_def}	
Z_{GC}\left({\mu_B\over T}, T, V\right)
=\sum_{j} \langle j |\exp\left({-\;\hat H + \mu_B \hat B \over T}\right) | j\rangle \;, 
\end{equation}
where $|j\rangle$ - eigenvectors of the Hamiltonian. In the case when summation is carried out only over $|j\rangle:\hat B |j\rangle = n |j\rangle$, we obtain the expression for the canonical 
partition functions: 
\begin{equation}\label{eq:ZC_def}	
Z_{C}\left(n, T, V\right)
=\sum_{j:\hat B |j\rangle = n |j\rangle } \langle j |\exp\left({-\;\hat H \over T}\right) | j\rangle \;.
\end{equation}

The corresponding fugacity expansion has a form:
\begin{equation}\label{eq:ZGC_fugacity_expansion}	
Z_{GC}(\theta, T, V)\;=\; \sum_{n=-\infty}^\infty Z_C(n,T,V)\xi^n,
\end{equation}
where $\theta = \mu_B/T =\theta_R + \imath\theta_I$ and    $\xi=e^{\theta}$ is fugacity.
We set the number of colors and the number of flavors to be equal to one, since the problem of negative probabilities arises even in this simplest case. In this case, the baryon number coincides with the fermion number --- in our work these are interchangeable terms. The symbols $T,V$ in partition function arguments will be systematically omitted.

From the relation (\ref{eq:ZGC_def}) it follows that $Z_{GC}$ is a periodic function of $\theta_I$: $Z_{GC}(\theta) = Z_{GC}(\theta+2\pi i)$. The inversion of the fugacity expansion has the form:
\begin{eqnarray}\label{Fourier}
Z_C\left(n\right)=\left. \int_{-\pi}^{\pi}\frac{d\theta_I}{2\pi}
e^{-in\theta_I}Z_{GC}(\theta)\right|_{\theta_\mathrm{R}=0}.
\end{eqnarray}

If a system of free fermions is characterized by 
temperature $T$ and the dimensionless baryon chemical potential $\theta$ then the probability that the baryon number of the system equals $n$ is given by the formula
\beqn
P_\theta (n)={Z_C(n)\xi^n \over Z_{GC}(\theta)}\,,
\eeqn
or, at $\theta=0$,
\beqn
P_n \equiv P_0(n)={Z_C(n) \over Z_{GC}(0)} \,.
\eeqn

The grand canonical partition function of free massless fermion gas at temperature $T$ in a box with edge length $L$ and volume $V = L^3$ has the form
\begin{equation} 
Z_{GC}(\theta)=\prod_{\mathbf{k}\in Z\!\!\!Z^3} 
\big( 1+ \xi w(\mathbf{k}) \big)^2 \; 
\big( 1+ \xi^{-1} w(\mathbf{k}) \big)^2\;,
\end{equation} 
where the wave vector $\mathbf{k}$ is quantized as $\mathbf{k} = 2\pi n_i / L$ with $n_i \in \mathbb{Z}$, 
\beqn
w(\mathbf{k})= \exp\left(\;-\;{E_{\mathbf{k}}\over T}\right), 
\eeqn
and 
\beqn
E_{\mathbf{k}} = {2\pi |\mathbf{k}| \over L}, \ \mathrm{and}\ {E_{\mathbf{k}}\over T} = {2\pi |\mathbf{k}| \over \sqrt[3]{\nu}};
\eeqn
where $E_{\mathbf{k}}$ is the energy and $\nu = L^3T^3 = VT^3$ 
is the dimensionless parameter characterizing the system under study;
its physical meaning is as follows: $\dfrac{\nu}{(2\pi)^3}$ 
is an estimate of the number of fermionic modes 
in a box of size $L$ excited at temperature $T$ or, 
on the other hand, this is the number of nodes of temperature waves 
in a box of volume $V$. Thus, the pressure can be represented as follows:
\beqn
p(\theta) & = & \frac{T}{V} \ln Z_{GC}(\theta) \nonumber\\
& = & \frac{2T}{V} \sum_{\mathbf{k}\in Z\!\!\!Z^3} \ln\big( 1+ e^\theta  w(\mathbf{k}) \big) \; + \Big( \theta \leftrightarrow -\;\theta\Big).
\eeqn
Replacing summation by integration according to the standard rules
\beqn
\sum_{\mathbf{k} \in \mathbb{Z}^3} ... \to \frac{VT^3}{(2\pi)^3} \int ... d\mathbf{q} 
\to \frac{VT^3}{2\pi^2} \int q^2\;dq ... \,,
\eeqn
changing the variables
\beqn
\mathbf{k} \to  \frac{ \sqrt[3]{\nu} \mathbf{q}}{2\pi}\;,
\eeqn
and applying integration by parts, we obtain
\begin{eqnarray}\label{eq:p_infty}
p_\infty(\theta) &=& \frac{T^4}{\pi^2} \int q^2\;dq \ln \left( 1+e^{\theta-q}\right) + \Big( \theta\leftrightarrow -\;\theta \Big) \\ 
&=& -\frac{2 T^4\;}{\pi^2} \left( \mathrm{Li}_{4}\big(-e^{-\theta}) + \mathrm{Li}_{4}\big(-e^{\theta}) \right)  \;,
\label{eq:polylog}
\end{eqnarray}
where symbol $\infty$ indicates that the pressure is obtained as a result of the transition to the infinite-volume limit by replacing summation with integration\footnote{In addition to $\infty$, the symbols $T$ (true) and $A$ will be used to indicate\-the quantities , calculated by definition and by asymptotic formulas, respectively.}. Using the identities for polylogarithms~\cite{ErdelyiA.1981} gives~\cite{Greiner:1995,Cleymans1986}:
\beqn
\hat p_\infty(\theta) & = & \displaystyle {1\over 6}
\left( {7\pi^2\over 30} + \theta^2 + {\theta^4\over 2\pi^2} \right)\,,
\ \mathrm{if}\  -\pi< \theta_I \leq \pi\;, \\
\hat p(\theta) & = & \hat p(\theta-2\imath \pi n),  
\ \mathrm{if}\  -\pi +2\pi n <\theta_I \leq  \pi +2\pi n \;, \nonumber
\label{eq:pressure_free}
\eeqn
where $\displaystyle \hat p = \frac{p}{T^4}$ is dimensionless pressure. 
Representing the grand canonical partition function in the form
\begin{equation}\label{eq:Z=e^vp}
Z_{GC}(\theta)=\exp \left[ \nu \hat p (\theta) \right]\; ,
\end{equation}
canonical partition functions can be written as integrals~\eq{Fourier}.

The periodic continuation of $Z_{GC}(\theta)$ from the strip $-\pi < \theta_I \leq \pi$ to the entire complex plane $\theta$ is a discontinuous function, since
\beqn
\label{eq:ZGC_strip}
Z_{GC}(\theta_R\pm i \pi) =  \exp\bigg[ {\nu\over 12\pi^2} \bigg( & - & \frac{8\pi^4}{15} + \theta_R^4-4(\theta_R)^2 \pi^2 \nonumber\\
& \pm &  4\imath \pi (\theta_R)^3 \bigg) \bigg]. 
\eeqn
Note that the discontinuity occurs at $\theta_R \neq 0$; in the three-dimensional case $Z_{GC}(\imath\theta_I)$ is a continuous function $\theta_I$. This discontinuity is associated with a cut along the negative imaginary semi-axis in the complex fugacity plane and, therefore, the fugacity expansion~\eq{eq:ZGC_fugacity_expansion} (that is, the Laurent series for $Z_{GC}(\xi)$) diverges everywhere except the circle $|\xi|=1$.

The grand canonical partition function for $\theta_R=0$ has the form:
\begin{equation}\label{eq:ZI_vs_thetaI}
Z_{GC}(i\theta_I) = A\exp \left[ \hat \Omega(\theta_I) \right]\; \quad \mbox{if} \quad -\pi< \theta_I \leq \pi\;,
\end{equation}
where
\begin{equation}\label{eq:Omega_def}
A=\exp\left(7\pi^2\nu\over 180 \right) \quad \mbox{and} \quad \hat \Omega(\theta_I)={\nu\over 6}\left( -\,(\theta_I)^2 + {(\theta_I)^4\over 2\pi^2}\right)\;.
\end{equation}
It continues beyond the interval $\theta_I\in (-\pi,\pi]$ according to the periodicity condition: $Z(\imath \theta_I)=Z(\imath \theta_I+ 2\imath \pi n)$. Although the functions $Z_{GC}(\imath\theta_I) = A e^{\hat \Omega} $ and $\hat \Omega(\theta_I)$ are continuous with two derivatives, the third derivative has a discontinuity at the points $\theta_I = (2n+1) \pi, \quad n\in Z\!\!\!Z$, which can be found as follows:

\begin{figure}[hht]
\centering
\includegraphics[width=87mm]{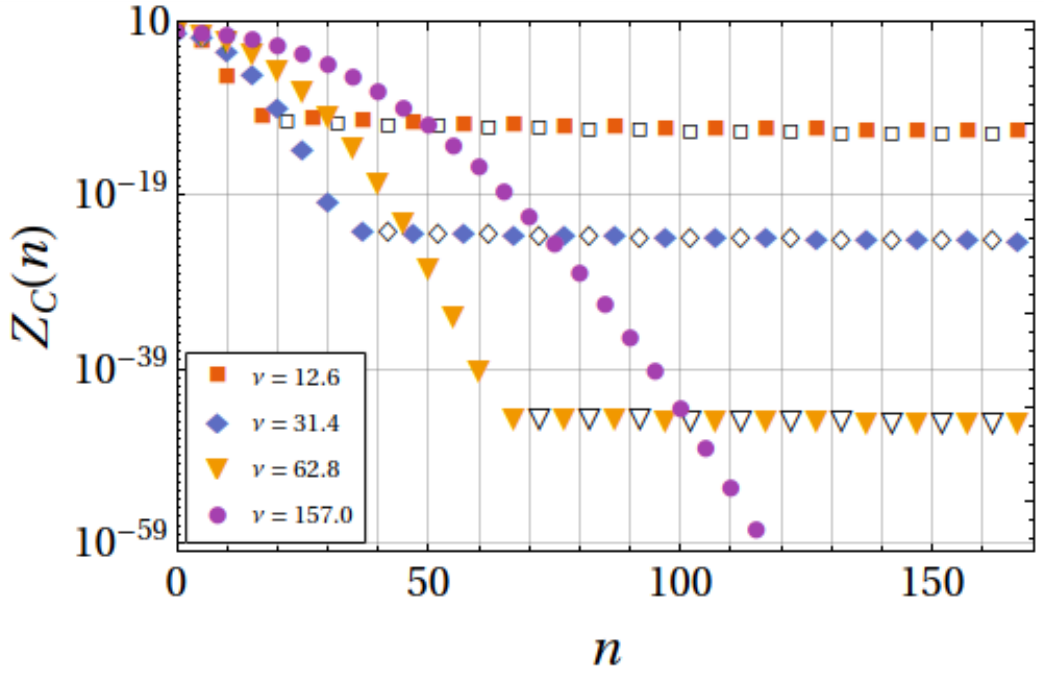}
\vspace{-3mm}
\caption{Canonical partition functions calculated using formulas (\ref{Fourier}) and (\ref{eq:Z=e^vp}) from pressure (\ref{eq:pressure_free}). The behavior at $n\gtrsim \nu $ is given by the formula (\ref{eq:alt_sign_asympt}). Blank symbols indicate absolute values of negative values.} 
\label{fig:Zn_neg}
\vspace{-5mm}
\end{figure}

\begin{eqnarray}
\left(e^{\hat \Omega}\right)'''\Big|^{\theta_I=\;\pi+0}_{\theta_I=\;\pi-0}
& = & e^{\hat \Omega}\Big|_{\theta_I=\pi} \cdot
\hat \Omega'''\Big|^{\theta_I=\;\pi+0}_{\theta_I=\;\pi-0} \nonumber\\
& = & 
- \exp\left( - \frac{\nu \pi^2}{12} \right) \frac{4\nu}{\pi}. \nonumber 
\end{eqnarray}
Thus, the function $\displaystyle \hat \Omega(\theta_I) \exp\left( -\;{\pi^2 \nu\over 12}\right)$ has exactly the same discontinuities as the function $\exp\big(\hat \Omega(\theta_I)\big)$. This fact  allows us to find the leading asymptotics  of the Fourier coefficients of the function (\ref{eq:Omega_def}) for $n\to\infty$, since it is completely determined by the discontinuities of the function and to find the asymptotic behavior of the canonical partition functions it is sufficient to find the Fourier coefficients of the function $ \hat \Omega(\theta_I)$: 
\begin{equation}
\hat \Omega(\theta_I)=
{4\nu \over\pi^2 } \sum_{n=1}^\infty {(-1)^n\over n^4} \Big( 1-\cos(n\theta_I)\Big)\;.
\end{equation}
The canonical partition functions for $n\to \infty$ take the form:
\begin{equation}\label{eq:alt_sign_asympt}
Z_C^\infty(n)= {2\nu\over \pi^2} \exp\left( \,-\,{2\pi^2\nu\over 45}\right) {(-1)^{n+1}\over n^4} + \underline{O}\left(1\over n^6\right)\;,
\end{equation}
the symbol $Z_C^\infty(n)$ means that these statistical sums were obtained using the formulas (\ref{Fourier}) and (\ref{eq:Z=e^vp}) with $\hat p=\hat p_\infty (\theta)$. Their behavior is shown in Fig.~\ref{fig:Zn_neg} and it is clearly seen that, at a certain critical value of $n$ depending on $\nu$, the sign of $Z_C^\infty(n)$ begins to alternate.

From the probabilistic point of view, $Z_{GC}(\imath\theta_I)/Z_{GC}(0)$ is the characteristic function of the distribution $P_n$. The obtained result with negative probabilities is a manifestation of a more general situation: according to Marcinkiewicz’s theorem, a polynomial of degree greater than two cannot be a characteristic function of a random variable \cite{Lukacs1970}. Our reasoning can be considered as a proof/extension of Marcinkiewicz’s theorem for some discrete random variables.

\section{One-dimensional gas of free massless fermions}
\label{sec:1D}

Partition function $Z^T_{GC}(\theta)$ 
of free massless fermions on a segment of length $L$ is given by the expression
\begin{equation} \label{eq:ZGC_true}
Z^T_{GC}(\theta) =
\prod_{n=1}^\infty (1+\xi w^n)^2(1+\xi^{-1}w^n)^2\,,
\end{equation}
$\displaystyle w=\exp\left(-\,\frac{2\pi}{\nu}\right)$ and $\nu=LT$.
Corresponding canonical partition functions
can be found from the expansion
$\displaystyle Z_{GC}(\theta)=\sum_{n=-\infty}^{\infty}Z_C(n)e^{n\theta} \;.$
The pressure corresponding to (\ref{eq:ZGC_true}) has the form \cite{abramowitz1972}:
\begin{equation}\label{eq:p_T_ultim}
\nu\hat p_T(\theta) = {\pi\over 2\nu} + 2 \ln \vartheta_{2}\!\left({i \theta\over 2\pi}; { i \over \nu}\right)
-2\ln\!\left[2\cosh {\theta\over 2}\right] -2\ln\phi(w),
\end{equation}
where
\begin{equation}
\vartheta_{2}(z; \tau ) = 2\sum_{n=0}^\infty \exp\left(i\pi\tau \left(n+{1\over 2}\right)^2\right) \cos \big((2n+1)\pi z \big),
\end{equation}
and $\phi(x)=\prod_{n=1}^\infty \big(1-x^n\big)$ is the Euler function. The corresponding canonical partition functions are well approximated by the formula
\begin{eqnarray}
 Z_C^{T,approx}(n) \simeq 
 {\displaystyle \exp\left({\pi\nu\over 6} + C(\nu) - {\pi n^2\over 2\nu}\right) \over\sqrt{2\pi\nu} \displaystyle \left( 1+\cosh\left( {\pi n\over \nu} \right)\right)} 
 \;.
\end{eqnarray}
where we adjust the constant $C(\nu)$ on the basis of numerical estimates.
At $C(\nu)=0$ and $\nu \geq 60$, the relative deviation 
$$\displaystyle R(n)={\ln Z_C^{T,approx}(n) - \ln Z_C^{T}(n) \over \ln Z_C^{T}(n)} $$ 
is rather small: $R(0)<3\cdot 10^{-3}$ and $R(n)$
does not exceed $2\cdot 10^{-5}$ at $n>500$. 
In so doing, the constant $C(\nu)$ can be adjusted so that $R(0)< 10^{-3}$ and $R(n>500)< 10^{-15}$.

The calculation of the logarithm of the partition function 
(\ref{eq:ZGC_true}) by the procedure similar to 
that described above gives the pressure 
in the infinite-volume limit as follows:
\begin{equation}
\hat p_\infty(\theta)=\frac{\theta^2}{2\pi} + \frac{\pi}{6}, \qquad \mbox{if}\quad |\theta_I|<\pi ,
\end{equation}
and then it is periodically continued to the remaining part of the complex plane using the condition $\hat p_\infty(\theta + 2\pi i ) = \hat p_\infty(\theta)$.
The corresponding canonical partition functions have the form
\begin{eqnarray}\label{eq:ZC_on_Segment}
Z_C^\infty(n)  &=& {1 \over 2\pi \nu} \exp\left( \frac{\nu\pi}{6}-\frac{\pi n^2}{2 \nu}\right)  \nonumber\\
&\times& \mathrm{Erf}\Big[\sqrt{\frac{\pi}{2\nu}} (i n - \nu),\;\sqrt{\frac{\pi}{2\nu}}(i n + \nu )\Big] \;, 
\end{eqnarray}
where $\mathrm{Erf}(z_0, z_1) = \mathrm{Erf}(z_1) - \mathrm{Erf}(z_0)$, and $\mathrm{Erf}(z)$ - is the error function,
$Z_C^\infty(n)$ have alternating sign at $n\gtrsim \nu$.
Canonical partition functions are obtained 
by the method of steepest descent in the limit 
$\nu\to\infty $. In the leading order they have the form
\begin{equation}
Z_C^A(n) \simeq {1\over \sqrt{2\nu}} \exp\left({\nu\pi\over 6} -\frac{\pi n^2}{2\nu}\right),
\end{equation}
and the respective pressure is as follows:
\begin{equation}\label{eq:p_A_ultim}
\nu \hat p_A = {\nu\pi \over 6} -{1\over 2}\ln 2\nu + \vartheta_3 \left( -\; {i \theta \over 2\pi }; {i \over 2\nu} \right),
\end{equation}
\begin{equation}
\mbox{where} \qquad \vartheta_3(z;\tau)\;=\; \sum_{n=-\infty}^\infty \exp\left(\pi i n^2 \tau + 2\pi i n z\right).
\end{equation}
\begin{figure}[hbt]
\centering
\includegraphics[scale=0.6]{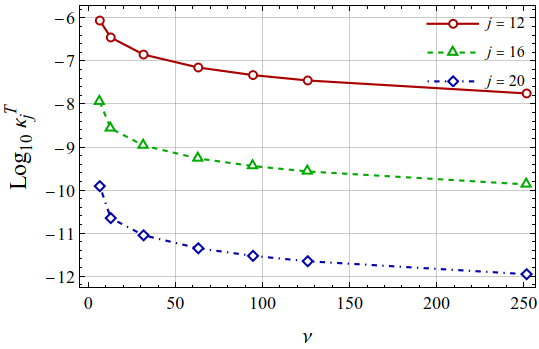}
\hspace{4mm}
\includegraphics[scale=0.6]{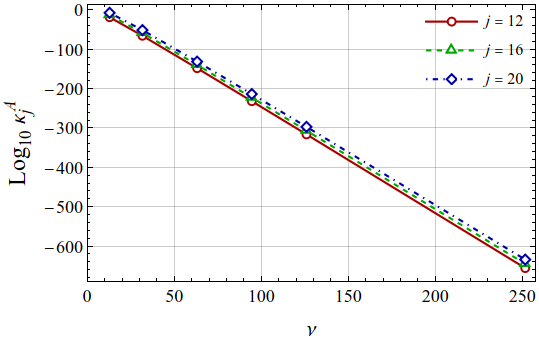}
\vspace{-3mm}
\caption{Cumulants $\kappa_j$ of the distribution ${\cal P}_n$ in
the net-baryon number for the one-dimensional case. Asymptotic 
cumulants increase with $n$, in contrast to true ones.} 
\label{fig:cum_high_1D}
\vspace{-5mm}
\end{figure}

The cumulants $\displaystyle \kappa_j=\nu {\partial^j \hat p\over \partial \theta^j}\Big|_{\theta=0}$ of the distribution $P_n$ show (Fig.~\ref{fig:cum_high_1D}) that the difference between the pressures $p_A$, $p_T$, and $p_\infty$ are rather small.

It should be noticed that, at a fixed value of $\nu$, 
the cumulants $\kappa_j^T$ decrease with $j$,
whereas the cumulants $\kappa_j^A$  increase.
Thus a naive extrapolation of our numerical results 
suggests that, at some $j=j_c$, $\kappa_j^A$ becomes equal to $\kappa_j^T$ and, at $j>j_c$, $\kappa_j^A$ exceeds $\kappa_j^T$.
An extrapolation based on Fig.~\ref{fig:cum_high_1D} suggests that 
$j_c$ is on the order of several hundred. Thus, we are tempted to 
study the details of the $\theta$-dependence of $p_A$ and $p_T$
in more detail.

\begin{figure*}[hbt]
\centering
\vspace{-28mm}
\includegraphics[width=95mm]{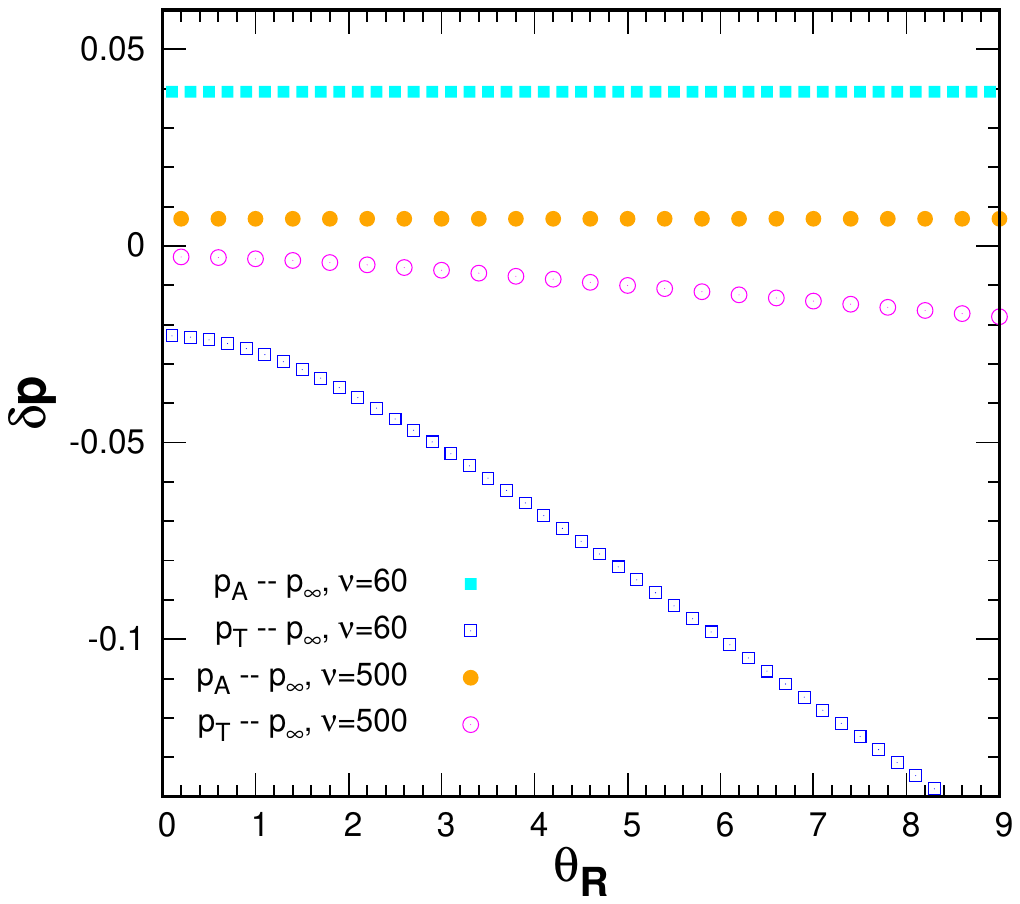}\hspace*{-7mm}\includegraphics[width=95mm]{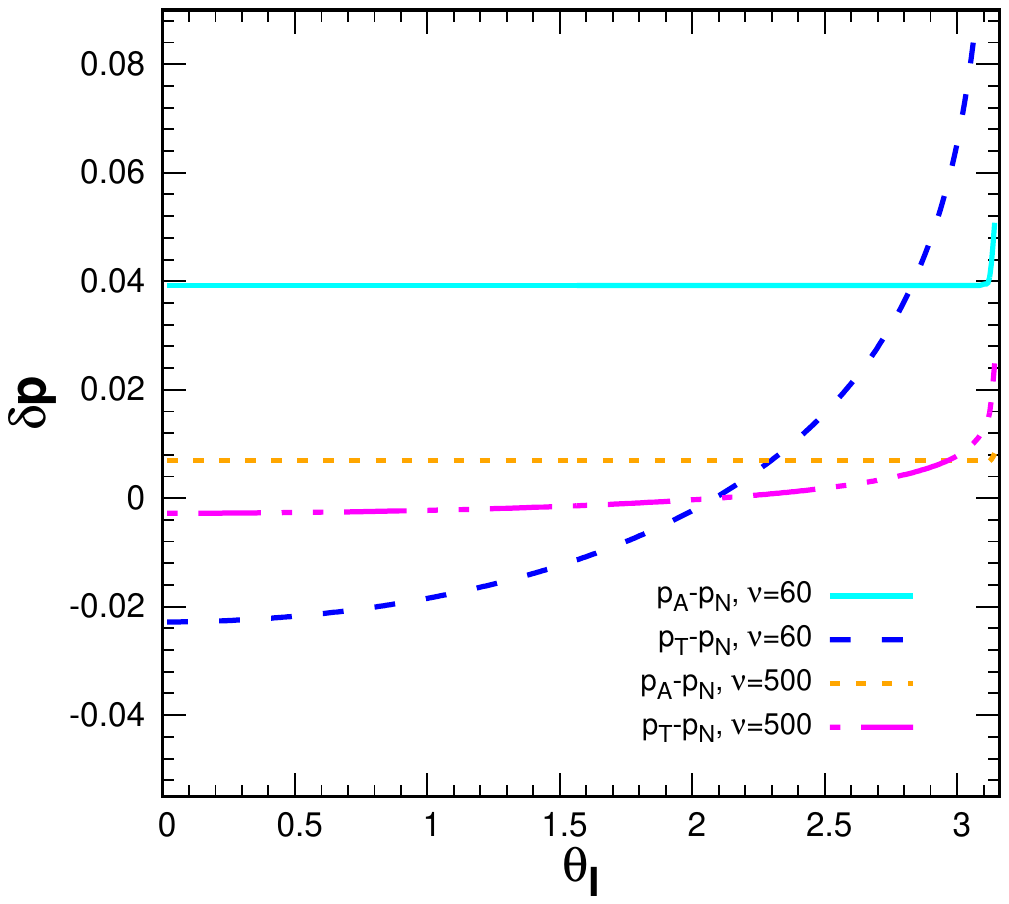}
\vspace{-35mm}
\caption{Differences $p_T-p_\infty$ 
and $p_A-p_\infty$ are plotted for $\theta_I=0$ (left panel) 
and for $\theta_R=0$ (right panel) at $\nu=60$ and 500.} 
\label{fig:deltap}
\vspace{-5mm}
\end{figure*}

The differences $p_A(\theta) - p_\infty(\theta)$ and 
$p_T(\theta) - p_\infty(\theta)$ are shown in Fig.~\ref{fig:deltap}. 

\begin{figure*}[hbt]
\centering
\vspace{-23mm}
\includegraphics[width=95mm]{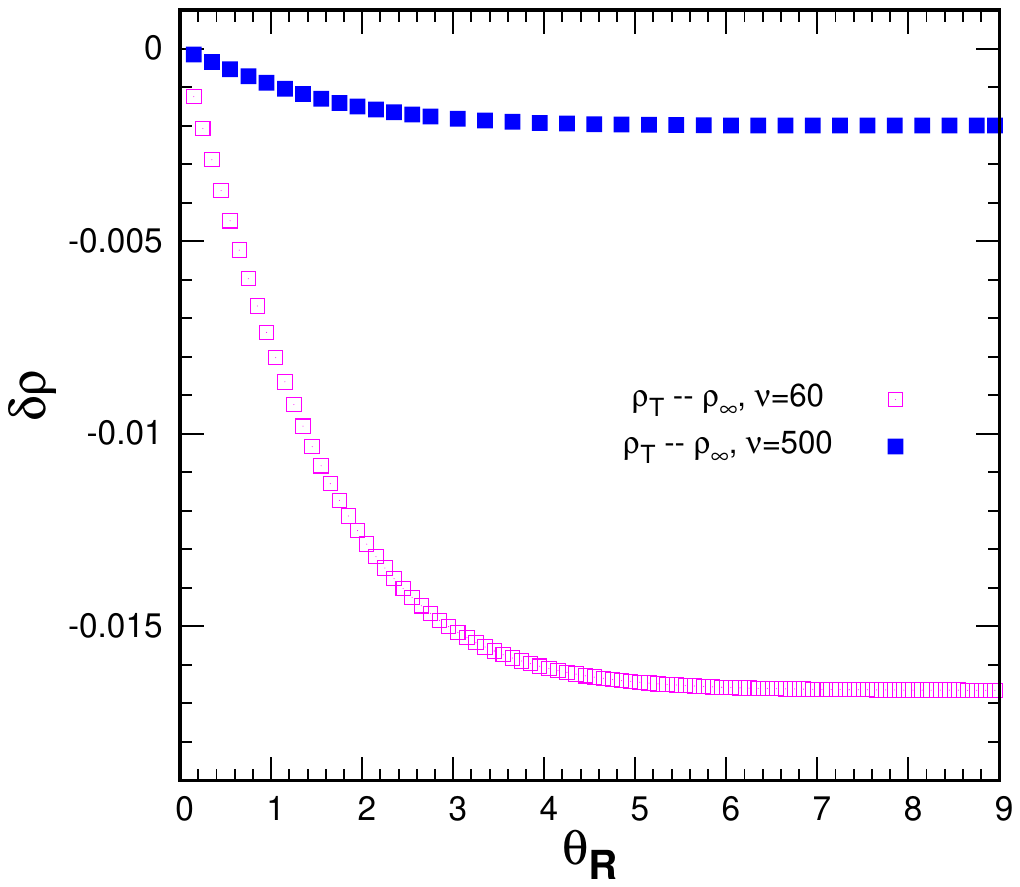}\hspace*{-7mm}\includegraphics[width=95mm]{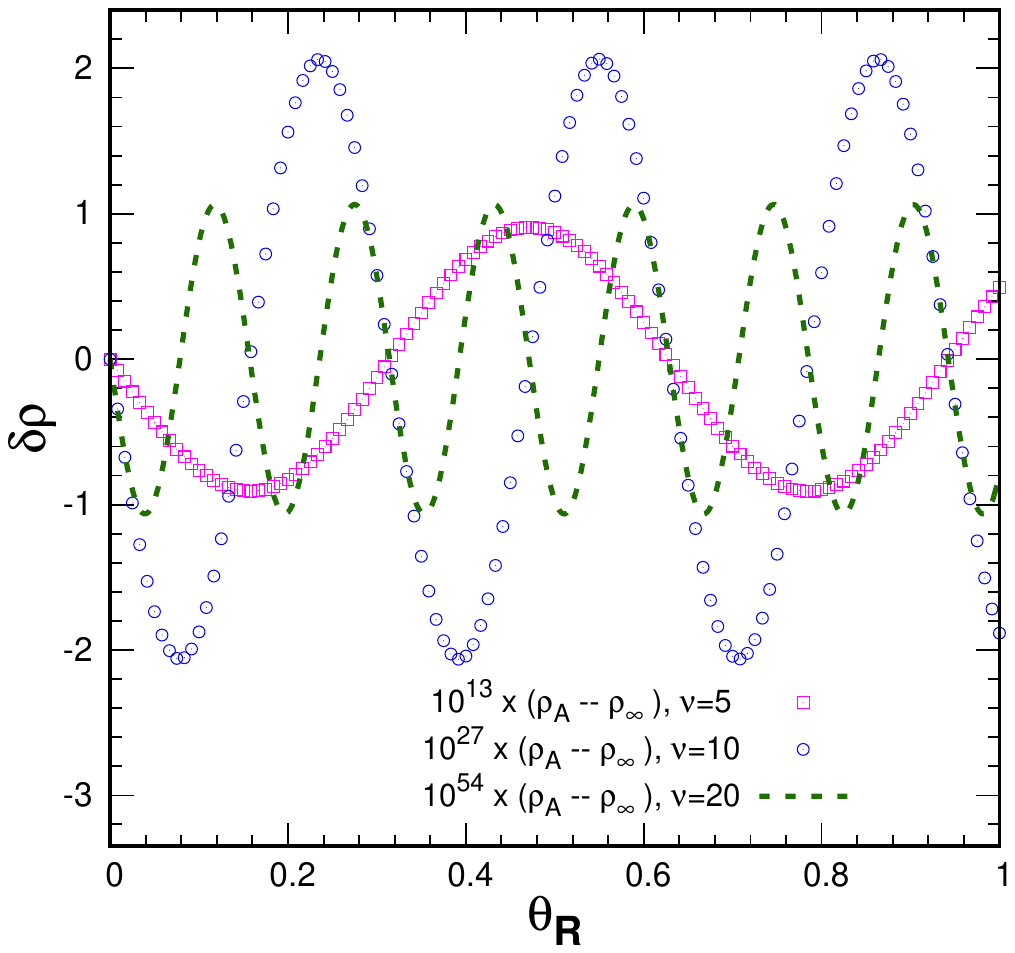}
\vspace{-35mm}
\caption{Differences 
$\hat \rho_T-\hat \rho_\infty$ (left panel)  and  
$\hat\rho_A-\hat \rho_\infty$ (right panel)
are plotted for and $\theta_I=0$ at various values of $\nu$.} 
\label{fig:rho_vs_thetaR}
\vspace{-5mm}
\end{figure*}

It is clearly seen that these differences decrease 
and come close to zero as $\nu$ increases, 
however, $p_A(\theta)-p_\infty(\theta)$ depends only weakly on $\theta$ and some features of its behavior may be poorly seen 
against the background of the constant term.
The distinctions between the asymptotic and true solutions  
are the most easily extracted from the $\theta$-dependence
of the respective baryon densities rather than 
from the pressures themselves. 
The differences $\hat \rho_T(\theta) - \hat \rho_\infty(\theta)$ 
and $\hat \rho_A(\theta) - \hat \rho_\infty(\theta)$ 
at real values of $\theta$ are plotted  
in Fig.~\ref{fig:rho_vs_thetaR}. 
Our numerical analysis reveals that 
the former difference tends to some constant value $\delta_T(\infty)$ at $\theta_R\to \infty$ 
(this limiting value is rapidly approached at $\theta_R>3$). 
This being so,
$\delta_T(\infty)$ decreases as $\displaystyle \sim {1\over \nu}$ at $\nu\to \infty$. The latter difference $\delta_A=\hat\rho_A(\theta) - \hat\rho_\infty(\theta)$ oscillates 
as $\displaystyle \delta_A\sim -A\sin (\omega\theta_R) $
with $\displaystyle A\sim \exp(-2\pi\nu)$ and $\omega\simeq 2\nu$
at $\nu\to\infty$ giving rise to the behavior of the highest-order cumulants $\kappa_j^{A}\sim \exp(-2\pi\nu) (2\nu)^{j}$,
in agreement with the results presented 
on the lower panel of Fig.~\ref{fig:cum_high_1D}.

\begin{figure}[hht]
\centering
\vspace{-23mm}
\includegraphics[width=95mm]{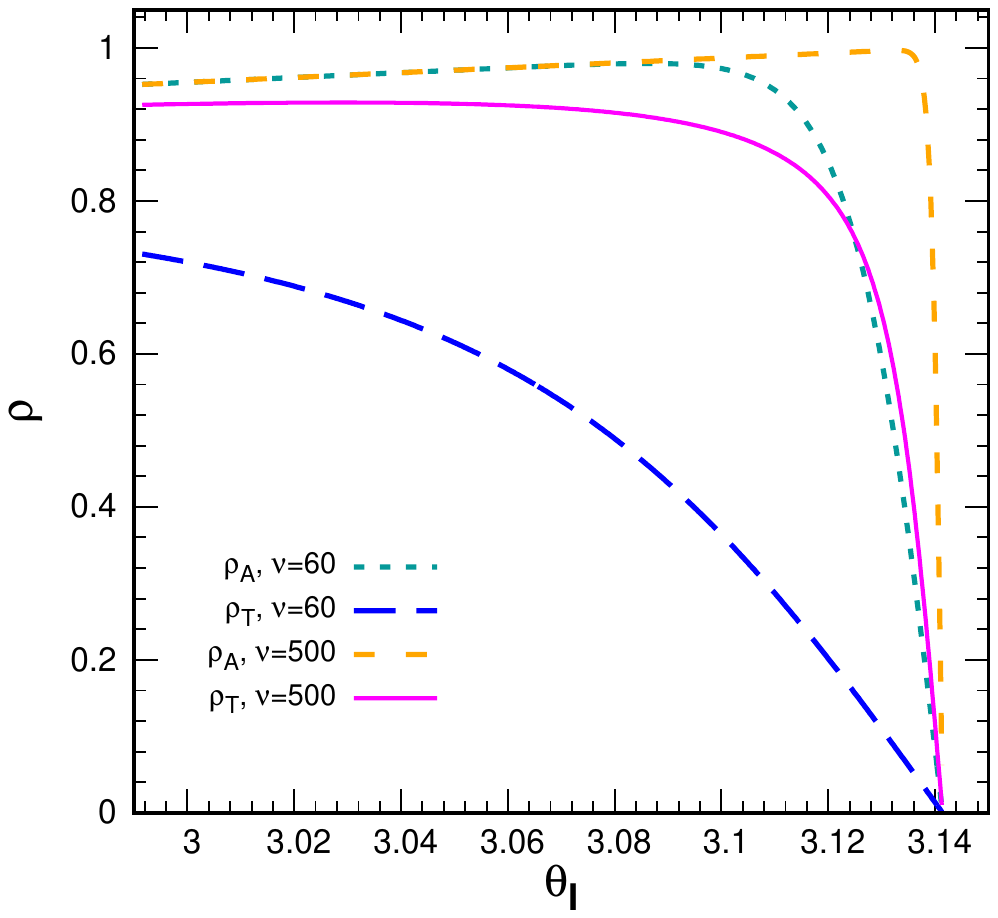}\vspace{-35mm}
\caption{The behavior of the pressures 
$\hat \rho_T(\theta_I)$  and  
$\hat\rho_A(\theta_I)$ in the neighborhood of the point $\theta=\imath \pi $ at various values of $\nu$.} 
\label{fig:neigh_of_ipi}
\vspace{-5mm}
\end{figure}

Therewith, it should be noted that ``asymptotic'' and ``true'' probabilities are significantly different for $|n|\gg \nu$:

\begin{equation} 
{\cal P}_n^A\sim {\cal P}_n^T \cosh\left( {\pi n\over \nu} \right)\;. 
\end{equation}

The authors of \cite{Roberge:1986mm,Rogalev_2023} proposed to use $\hat p_A(\theta)$ instead of $\hat p_T(\theta)$ for an approximate calculation of $Z_C(n)$, in spite of the fact that the pressure $\hat p_\infty(\theta)$, which is very close to $\hat p_A(\theta)$, is unphysical. Our example shows that corrections to the probability distributions associated with small variations of the pressure can be significant, despite their relative smallness.

\section{Roberge-Weiss transition}
\label{sec:RW}

The final formulas for pressure $\hat p_A(\theta,\nu)$ (\ref{eq:p_A_ultim})
and $\hat p_T(\theta,\nu)$ (\ref{eq:p_T_ultim}) represent analytical and $2\pi$-periodic in $\theta_I$  functions that converge at $\nu\to\infty$ to $\hat p_\infty(\theta)$ (\ref{eq:pressure_free}). 
The convergence of the respective baryon densities 
$\hat \rho_A(\theta_I)$ and $\hat \rho_T(\theta_I)$
to the function $\displaystyle \hat \rho_\infty(\theta) ={\partial \hat p_\infty(\theta) \over \partial \theta}\;$,
which is a discontinuous at $\theta_I=(2n+1)\pi,\ n\in \mathbb{Z},$
is illustrated in Fig.~\ref{fig:neigh_of_ipi}.
It is seen that $p_A(\theta_I)$ converges faster than
$p_T(\theta_I)$.

In the complex $\theta$ plane, the density $\displaystyle \hat \rho_\infty(\theta) $ is discontinuous on the lines $\theta_I=2\pi n \ (n \in \mathbb{Z})$,
where the zeros of the polynomials 
\begin{equation}
\sum_{n=-N}^N Z_C^A(n) \xi^{n+N}\quad\mbox{and}\quad \sum_{n=-N}^N Z_C^A(n) \xi^{n+N}
\end{equation} 
are located (these lines correspond to the real negative semiaxis in the $\xi=e^\theta$ plane). The distance between adjacent zeros is $\displaystyle \sim \nu^{-1}$ and, therefore, their density increases indefinitely as $\nu\to \infty$.

Thus, the Lee-Yang zeros occupy the negative real semi-axis in the fugacity plane. Provided that their density does not vanish
at $\xi=-1$ (this condition is fulfilled in the one-dimensional case), this fact indicates the first-order Roberge-Weiss phase transition in the one dimensional free theory. It should be emphasized that such a transition is not associated with strong interactions, in contrast to Roberge-Weiss's ideas \cite{Roberge:1986mm}.
 
In the three-dimensional case, integral estimation (\ref{Fourier})
by the saddle-point method gives $Z_C(n)\approx Z_C^A(n)$,
which are positive for sufficiently large $\nu$:
\begin{equation}\label{eq:asympt_main}
Z_C^A(n)=\exp\left( -\nu f\left({n\over \nu}\right)\right)\;.
\end{equation}
In the leading order in $\nu$, we arrive at 
\begin{eqnarray}\label{eq:logZC_free}
f(\hat \rho) &=& {\pi^2\over 18} +{9\pi\sqrt{3}\over 8} \hat\rho
\big( q^{1/3} - q^{-1/3}\big) \\ 
&& - {\pi^2\over 36}
\sqrt{1+{243\over 4\pi^2} \hat\rho^2} \big( q^{1/3} + q^{-1/3} \big)\;, \nonumber 
\end{eqnarray}
where
$$
q^{\pm 1} = \sqrt{1+{243\over 4\pi^2} \hat\rho^2} \pm {9\sqrt{3}\over 2\pi} \hat\rho \;.
$$
For $n\gg\nu$ we have $\displaystyle Z_C(n=\nu\hat \rho)\sim \exp\left( -\;{\nu\,\sqrt[3]{81\pi^2 \hat\rho^4}\over 4}\right)$, as opposed to (\ref{eq:alt_sign_asympt}).
Substituting the coefficients (\ref{eq:asympt_main}) into the formula~\eq{eq:ZGC_fugacity_expansion}and taking into account the relation (\ref{eq:Z=e^vp}), we obtain the pressure $\hat p(\theta )$
in the form of an entire and $2\pi$-periodic in $\theta_I$ function $\hat p_A(\theta,\nu)$, which for $\nu\to\infty$ converges to $\hat p_\infty(\theta)$ (\ref{eq:pressure_free}).
The fermion number density in three dimensions is discontinuous at $\theta_R\neq 0$, whereas at $\theta_R=0$  it is continuous with its first derivative and its second derivative is discontinuous.
That is, the Roberge-Weiss transition at $\theta_R=0$ is of the third order, according to the Ehrenfest classification.

At the same time, the standard arguments used to justify the saddle-point method cannot be considered as an evidence for the validity of the estimates
(\ref{eq:asympt_main})-(\ref{eq:logZC_free})
at high densities ($n\gg\nu$ or $\hat \rho \gg 1$).
However, there are the following considerations.
In the limit $\nu\to \infty$, $n\to \infty$, $\displaystyle {n\over \nu}=\hat \rho =$const, the net-baryon number density
$\hat\rho$ can be considered as a continuous function, $P_n\to \exp\big(-\nu \hat F(\hat \rho)\big)\;$,
and the fugacity expansion is given as follows:
\begin{equation}\label{eq:ZGC_recast}
Z_{GC}(\theta) =\nu\!\int \! d\hat\rho \;\exp\Big(\nu \big( \hat\rho\theta - \hat F(\hat \rho ) \big)\Big)\;.
\end{equation}
The canonical partition functions $Z_C(n)\sim \exp(-\nu \bar F(\hat \rho))$, evaluated in the leading order of the saddle-point approximation, can be derived using the function $\bar F(x)$, obtained from pressure by the Legendre transformation:
$\bar F(x) = x \theta_s-\hat p_\infty(\theta_s)$
where $\theta_s(x)$ is determined from the equation
$\displaystyle {\partial \hat p_\infty \over \partial\theta} \Big|_{\theta=\theta_s}=x$.
However\cite{Huangs1987}, the result of the Legendre transformation is the function $\bar F(\bar \rho)$, defined parametrically by the formulas
\begin{eqnarray}
\bar F &=& \int d\hat\rho \;\hat F(\hat \rho )\;\exp\Big(\nu \big( \hat\rho\theta - \hat F(\hat \rho ) \big)\Big)\;, \\
\bar \rho &=& \int d\hat\rho \;\hat\rho\;\exp\Big(\nu \big( \hat\rho\theta - \hat F(\hat \rho ) \big)\Big)\;.
\end{eqnarray}
If at $\nu\to\infty$ the distribution ${\cal P}_n$ has the  variance $\displaystyle \nu^{-1}$, there is reason to assume that formal replacement of $\bar F(\bar \rho)$ with $\hat F(\hat\rho)$ leads to corrections of the order of $\displaystyle \nu^{-1}$, however, the question of the consequences of such a replacement requires more careful study, since ${\cal P}_n$ instability occurs
with respect to small variations of $\hat p(\theta)$.

\section{Conclusions}

We have carried out a critical analysis of traditional methods for calculating the pressure, density, and fermion number distribution of a free massless fermion gas at finite temperatures. 
The pressure $p_\infty(\theta)$ calculated by the conventional procedure, that is, by employing a naive transition to the infinite-volume limit, leads to incorrect results for the net-baryon number distribution. A precise numerical calculation of the probability mass function gives negative values of some probabilities.

For this reason, we have considered the correct transition to the infinite-volume limit for free massless fermions on a segment, when the pressure and density, as well as the associated fermion number distribution, are calculated in a finite volume, and only then does the volume tends to infinity.
In this case, we have found the formulas for the pressure $\hat p_T(\theta)$ and $\hat p_A(\theta)$, which differ from the polynomial $\hat p_\infty(\theta)$  by the functions 
vanishing in the infinite-volume limit, and these formulas give ${\cal P}_n>0\ \forall n$ whereas $\hat p_\infty (\theta)$ gives ${\cal P}_n<0$ for some $n\gtrsim \nu$.
This indicates an instability of the net-baryon number distribution with respect to small variations in pressure as a function of the chemical potential, which requires further study.

We have shown that the baryon density becomes a discontinuous function in the complex $\theta$ plane for $\theta_I = (2n+1)\pi, n\in \mathbb{Z}$ 
in the infinite-volume limit, and the Lee-Yang zeros are concentrated on the discontinuity line, which is mapped to 
the negative real semi-axis in the fugacity plane. It indicates the first-order Roberge-Weiss phase transition   for a gas of free massless fermions for both one-dimensional and three-dimensional cases at $\theta_R\neq 0$; for $\theta_R=0$ in three dimensions the Roberge-Weiss transition is of the third order. Thus, the Roberge-Weiss transition in QCD at $T>T_{RW}$ may be caused by the presence of free massless fermion states in the particle spectrum.

\acknowledgments
The work of NVG was supported
by Grant No. 23-12-00072 of the Russian Science Foundation.
The work of AAK was supported
by Grant No. FZNS-2024-0002 of the Ministry of Science and Higher Education of Russia.

\bibliography{free_partition}

\end{document}